\documentclass[printer,traditabstract]{aa}
\usepackage{graphicx}
\usepackage{txfonts}
\usepackage{natbib}

\def\HI{H{\,\small I}}

\newcommand{\kms}{$\,$km$\,$s$^{-1}$}

\def\HI{H{\,\small I}}

\def\emph#1{{\sl #1}}
\newcommand{\ltsima} {$\; \buildrel < \over \sim \;$}
\newcommand{\gtsima} {$\; \buildrel > \over \sim \;$}
\newcommand{\lta} {\lower.5ex\hbox{\ltsima}}
\newcommand{\gta} {\lower.5ex\hbox{\gtsima}}

\newcommand{\sauron}{{\texttt {SAURON}}}
\newcommand{\atlas}{{ATLAS$^{\rm 3D}$}}

\begin{document}

\title{Probing the gas content of radio galaxies \\
through HI absorption stacking}

\author{K. Ger\'{e}b $^{1,2}$, R. Morganti$^{1,2}$, T.A. Oosterloo$^{1,2}$}
\authorrunning{Ger\'{e}b et al.}

\institute{$^{1}$Netherlands Institute for Radio Astronomy (ASTRON), P.O. Box 2, 7990 AA Dwingeloo, The Netherlands \\ $^{2}$Kapteyn Astronomical Institute, University of Groningen, P.O. Box 800, 9700 AV Groningen, The Netherlands}


\abstract{
Using the Westerbork Synthesis Radio Telescope, we carried out shallow \HI\ absorption observations of a flux-selected (S$_{\rm{1.4 \ GHz}}$ $>$ 50 mJy) sample of 93 radio active galactic nuclei (AGN) which have available SDSS (Sloan Digital Sky Survey) redshifts between 0.02 $<$ z $<$ 0.23. Our main goal is to study the gas properties of radio sources down to S$_{\rm{1.4 \ GHz}}$ flux densities not systematically explored before using, for the first time, stacking of absorption spectra of extragalactic \HI.
Despite the shallow observations, we obtained a direct detection rate of $\sim$ 29$\%$, comparable with deeper studies of radio galaxies. Furthermore, detections are found at every S$_{\rm{1.4 \ GHz}}$ flux level, showing that \HI\ absorption detections are not biased toward brighter sources.
The stacked profiles of detections and non-detections reveal a clear dichotomy in the presence of \HI, with the 27 detections showing an average peak $\tau = 0.02$ corresponding to N(\HI) $\sim (7.4 \pm 0.2) \times 10^{18}$ (T$_{\rm{spin}}$/c$_{\rm{f}}$) cm$^{-2}$, while the 66 non-detections remain undetected upon stacking with a peak optical depth upper limit $\tau< 0.002$ corresponding to N(\HI) $< (2.26 \pm 0.06) \times 10^{17}$ (T$_{\rm{spin}}$/c$_{\rm{f}}$) cm$^{-2}$ (using a FWHM of 62 \kms, derived from the mean width of the detections). Separating the sample into compact and extended radio sources increases the detection rate, optical depth, and FWHM for the compact sample. The dichotomy for the stacked profiles of detections and non-detections still holds between these two groups of objects. 
We argue that orientation effects connected to a disk-like distribution of the \HI\ can be partly responsible for the dichotomy that we see in our sample. However, orientation effects alone cannot explain all the observational results, and some of our galaxies must be genuinely depleted of cold gas.
A fraction of the compact sources in the sample are confirmed by previous studies as likely young radio sources (compact steep spectrum and gigahertz peaked spectrum sources). These show an even higher detection rate of 55\%. Along with their high integrated optical depth and wider profile, this reinforces the idea that young radio AGN are embedded in a medium that is rich in atomic gas.
Part of our motivation is to probe for the presence of faint \HI\ outflows at low optical depth using stacking. However, the stacked profiles do not reveal any significant blueshifted wing. We are currently collecting more data to investigate the presence of outflows.
The results presented in this paper are particularly relevant for future surveys in two ways. The lack of bias toward bright sources is encouraging for the search for \HI\ in  sources with even lower radio fluxes planned by such surveys. The results also represent a reference point when searching for \HI\ absorption at higher redshifts.}

\maketitle

\section{Introduction}\label{Intro}

The nuclear activity in galaxies is regulated by the availability of gas in the central region and by the conditions that make it possible for the gas to cool and feed the central black hole. Our knowledge of the gas content in nearby early-type galaxies (the typical hosts of radio-loud AGN)
has increased substantially in recent years. Projects like WSRT-\sauron\ \citep{Morganti2006, Oosterloo2010} and \atlas\ \citep{Serra, Young2011, Davis2013} have studied the presence and properties of \HI\ and molecular gas in nearby early-type galaxies. For example, detailed studies of the \HI\ content of early-type galaxies in the nearby Universe has shown that \HI\ is detected in emission in about 40 percent of early-type galaxies in the field \citep{Serra}. However, because these samples are limited to very nearby galaxies, only a minority of the objects host an AGN and, in particular, a radio AGN.

In radio-loud AGN, the presence and kinematics of the gas can be explored via \HI\ absorption, and this has been done {for} a long time \citep[e.g., ][]{Roberts1970, deYoung1973, Dickey1982, Shostak1983}. Such observations of \HI\ absorption are typically done using strong radio sources as a bright background against which the gas can be traced. A number of such studies have recently provided a better understanding of the \HI\ absorption properties of radio AGN \citep{Gorkom, Morganti2001, Vermeulen, Gupta2006, Curran2010, Emonts}. These samples show a detection rate of \HI\  between a few percent up to 40$\%$. Absorption was found to trace a variety of structures showing that \HI\ can be present in regularly rotating disks \citep{Serra, Emonts, Gallimore, Allison2014}, in infalling clouds that have been associated with the feeding mechanisms of the central black hole \citep{Gorkom, Morganti2009} and in outflows tracing the interactions between the jets and the surrounding medium \citep{Morganti1998, Morganti2005, Morganti2013}. Thus, the complexity of the \HI\ kinematics in AGN suggests that gas can play many different roles.

One  interesting finding of these studies is that there appears to be a trend between the detection rate and the type of radio source and, in particular, its evolutionary stage.
Compact steep spectrum (CSS) and gigahertz peaked spectrum (GPS) sources have been proposed to represent young (\ltsima10$^{4}$ yr) radio AGN, based on spectral aging analysis and lobe expansion speed measurements \citep{Fanti1995, Readhead1996, Owsianik1998}. A high detection rate of \HI\ absorption in these sources has been reported  by a number of studies \citep{Pihlstrom, Gupta2006, Emonts, Chandola}.
This has been interpreted as evidence of a relation between the recent triggering of the AGN activity and the presence of \HI\ gas.

Finally, the presence of a strong interaction between the radio jets and the interstellar medium has been proposed for many of these young sources. Although it is not yet clear in how many cases this would result in the "frustration" of the radio source \citep{Breugel1984, Fanti1990, Young1993, Pihlstrom}. Outflows are often seen in the kinematics of the \HI\ in these objects, as traced by blueshifted, broad wings of absorption (see e.g., Morganti et al. 2013 and ref. therein). These examples show that, at least in some cases, the jets are clearing their way through the surrounding gas.

It is clear that a better understanding of the gas properties of different types of radio sources (e.g.,\ compact, extended) is crucial for explaining the observed characteristics of AGN. It is also of interest to investigate whether \HI\ has a similar detection rate and morphology in early-type galaxies with/without AGN and hence to learn more about the interplay between AGN and the surrounding gas.
Previous studies of \HI\ in AGN are limited in number and sensitivity, not ideally suited to study weak structures like outflows. Some of these limitations will be overcome by surveys performed with new telescopes, such as Apertif \citep{Oosterloo2010b}, the Australian Square Kilometre Array Pathfinder (ASKAP, DeBoer et al. 2009), and MeerKat \citep{Booth2009}. However, even with current telescopes progress can be made using stacking techniques. \HI\ emission stacking techniques have been used effectively to detect the global signal from faint \HI\ emitters \citep{Lah2007, Verheijen2007, Lah2009, Fabelloa, Fabellob, Delhaize, Gereb}. More recently, \cite{Murray2014} performed a spectral stacking analysis using \HI\ absorption residual spectra to detect warm neutral medium in the Milky Way. 

Here we present the results of a snap-shot survey we have performed with the main goal of observing as many objects as possible, to a continuum flux density limit that is lower than the earlier studies. For the first time, we stack extragalactic data of \HI\ absorption to lower the optical depth detection limit, and to study the gas properties of different types of AGN. In an upcoming paper we will present the data and the discussion of the single objects and their \HI\ profile parameters.

In this paper the standard cosmological model is used, with parameters $\Omega_{\Lambda}$ = 0.3, $\Lambda$ = 0.7 and $H_0$ = 70 km s$^{-1}$ Mpc$^{-1}$.

\section{Sample selection and observations}\label{Sec:SampleSelection}

For our sample selection, we used the cross-correlation of the Sloan Digital Sky Survey (SDSS, York et al. 2000) and Faint Images of the Radio Sky at Twenty-cm (FIRST, Becker et al. 1995) catalogs. In total, 120 sources were selected with peak flux $\rm{S_{1.4 \ GHz}}$ $>$ 50 mJy in the FIRST catalog, in the redshift range $0.02 < z < 0.23${\footnote{We start from redshift z = 0.02 in order to exclude nearby, well-studied galaxies.}, above the declination $\delta >15 \deg$. 
The interval $0.136 < z < 0.175$ was excluded because it is strongly affected by radio frequency interference (RFI). Additional information on the radio properties of the sample is extracted from the 1.4 GHz NRAO VLA Sky Survey (NVSS, \citealt{Condon1998}).

The SDSS spectra were visually inspected to ensure that accurate optical redshifts can be derived for the galaxies. For nine sources the spectra appear without any well-defined emission/absorption lines, meaning that for these sources the optical line identification, and hence the redshift in SDSS, may not be reliable. For this reason, we excluded these sources from our sample. For the rest of the sample, the redshifts errors extracted from SDSS are lower than $\sim$ 60 \kms.

\begin{figure}[t!]
\begin{center}
\includegraphics[width=.45\textwidth]{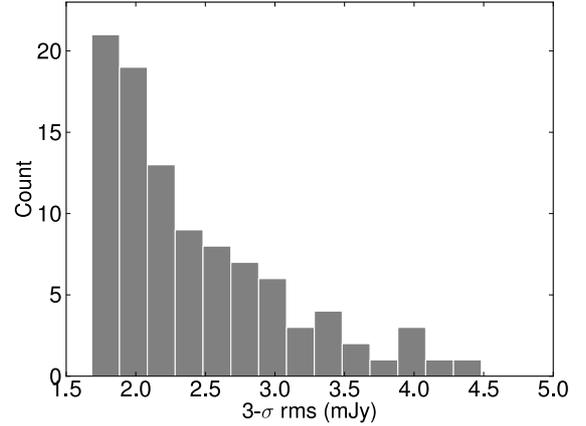}
\caption{The 3-$\sigma$ rms noise distribution in the 101 \HI\ data cubes.}\label{fig:Noise}
\end{center}
\end{figure}

\begin{figure}
\begin{center}
\includegraphics[width=.45\textwidth]{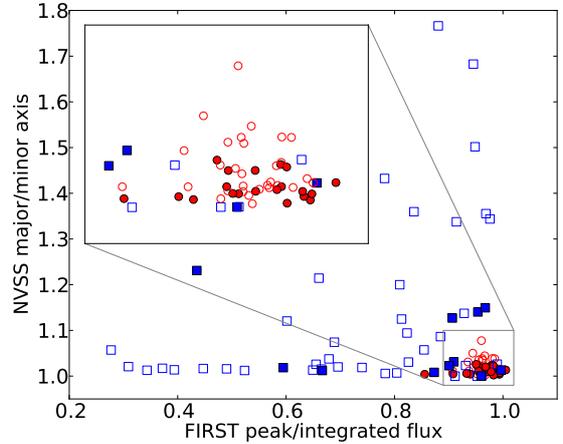}
\caption{Radio morphological classification of the observed sample (101 sources). Red circles indicate compact sources, and extended sources are marked by blue squares. \HI\ detections (both compact and extended) are marked by filled symbols, while empty markers stand for non-detections.}\label{RadioMorpho}
\end{center} 
\end{figure}

The observations were carried out with the Westerbork Synthesis Radio Telescope (WSRT) in the period Nov. 2012 - Nov. 2013. Using the WSRT East-West array, the short integration time of 4 hours for each source means that the synthesised beam of the data cubes is very elongated, with a typical angular size of 75 $\times$ 11 arcsec. The observational setup consists of 1024 channels in a bandwidth of 20 MHz. We discard 10 more objects because their spectra are dominated by RFI. As a result, our final AGN sample includes 101 sources. 

The data were reduced using the MIRIAD package \citep{Sault}. The \HI\ data cubes were Hanning smoothed over 3 channels, yielding a final velocity resolution of $\sim$ 16 km s$^{-1}$. The 3-$\sigma$ rms noise distribution of the \HI\ data cubes are presented in Fig. \ref{fig:Noise}.

We also created continuum images using the line-free channels to measure the continuum flux density of the sources. The corresponding radio power distribution ranges between 10$^{23}$ - 10$^{26}$ W Hz$^{-1}$.

\subsection{Characterization of the AGN sample}

The shortlisted target sample was separated into compact and extended radio sources. Our sample is small enough that this classification could be done visually. However, we also explored automated procedures based on objective parameters, which could be used in larger surveys of radio sources in the future. We found the best method to be the classification based on the NVSS major-to-minor axis ratio vs.\ the FIRST peak-to-integrated flux ratio. As illustrated in Fig. \ref{RadioMorpho}, and after a final visual inspection, we find that except for one source, all compact sources are located in the region with NVSS major-to-minor axis ratio $<$ 1.1 and FIRST peak-to-integrated flux ratio $>$ 0.9 (marked by the box). Most of the extended sources have NVSS major-to-minor axis ratio $> 1.1$ and FIRST peak-to-integrated flux ratio $<$ 0.9. 
However, even this method is not perfect and we identify a few more extended sources inside the box, which seem to be misclassified because these sources tend to show strong core emission compared to the faint extended emission coming from the lobes.
Furthermore, following a literature search of the detections we find that one source is classified as an extended source in our selection, however it is a 6 kpc Compact Steep Spectrum (CSS) source \citep{Saikia2003}. This object was misclassified because the structure of the radio source is slightly resolved at the 5 arcsec resolution of FIRST. Thus, this source is added to the sample of compact sources.

The observed sources are a mix of different types of AGN. Because our sample is solely flux selected, the sample contains radio galaxies (the majority of the sources), QSOs (Quasi-Stellar Objects), Seyfert galaxies, gas-rich mergers. Radio galaxies are typically found in red ($g - r$ $>$ 0.7) early-type hosts \citep{Bahcall, Best2005}, while the other three groups are typically blue ($g - r$ $<$ 0.7) objects and richer in \HI\ \citep{Gereb}.
In order to make the sample more homogeneous, we have excluded blue galaxies with $g - r$ $<$ 0.7 from our analysis. The color information needed for this exercise, i.e., the $g$ and $r$ band optical magnitudes, are extracted from the SDSS database. As a result of this selection, we exclude gas-rich mergers, QSO-s, Seyfert galaxies, such that the remaining sample of 93 AGN mainly consists of radio galaxies.  

\begin{figure}
\begin{center}
\includegraphics[width=.45\textwidth]{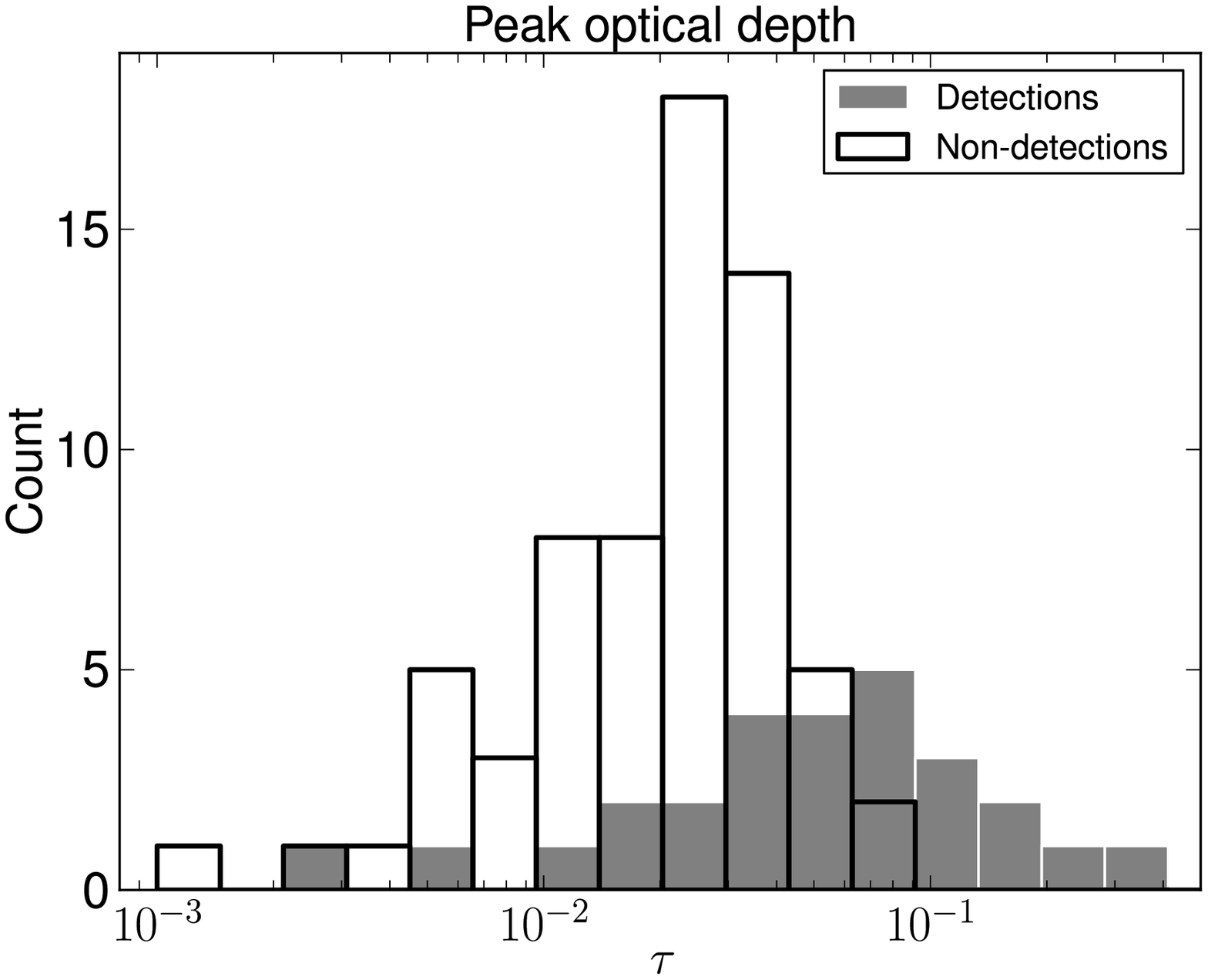}
\includegraphics[width=.45\textwidth]{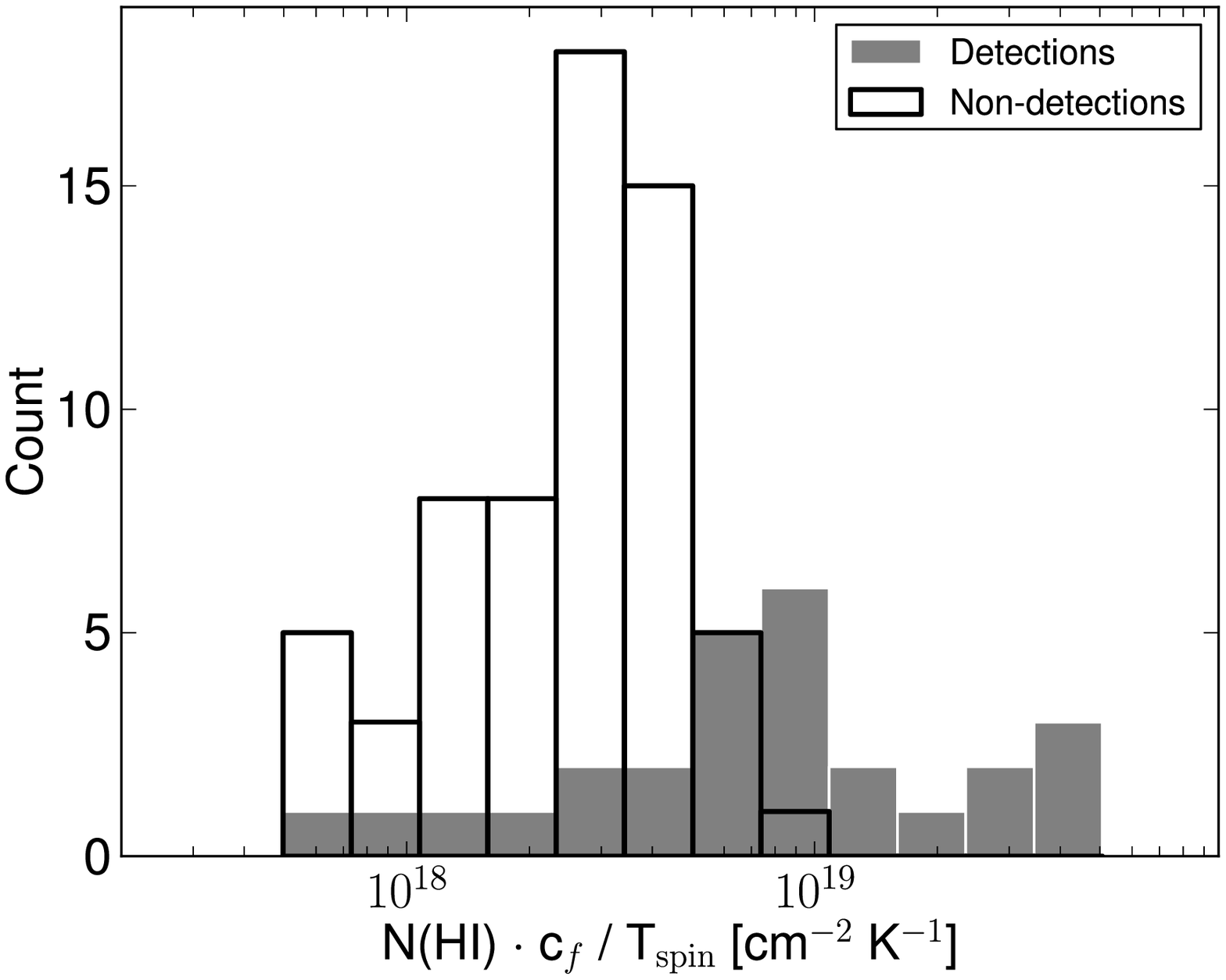}
\caption{Peak optical depth and N(\HI) distribution of the detections (filled bars) and 3-$\sigma$ upper limits (empty bars). }\label{fig:TauNHI}
\end{center} 
\end{figure}

\section{Results}\label{Results}

\subsection{\HI\ detection}\label{Detection_HI}

After the above selection, our sample contains 93 $g - r > 0.7$ AGN, of which we detected 27 sources above the 3-$\sigma$ level, and 66 were non-detections. A first result is therefore that, despite the short integration time, we obtained a high detection rate of $\sim 29\%$, comparable with deeper studies of radio galaxies. 

The absorbed flux depends both on the optical depth and on the covering factor (c$_{\rm{f}}$) of the \HI, such that \mbox{S$_{\rm abs}$ = $S \rm{c_f} (1 - e^{-\tau})$}. The column density (in cm$^{-2}$) is related to the integrated optical depth (in \kms) by \mbox{N(\HI) = 1.823$ \ \times \ 10^{18} ($T$_{\rm spin} / \rm{c_f} \int \tau(v)dv$} \citep{Wolfe1975}. As we mention in Sec.\ \ref{Sec:SampleSelection}, we derive the optical depth for each spectrum using the WSRT continuum flux. 
As illustrated in Fig. \ref{fig:TauNHI}, the detected $\tau$ distribution is very broad, with the peak $\tau$ of the direct detections ranging between 0.2$\%$ and 30$\%$. The corresponding column densities are N(\HI) $= 5 \times 10^{17}$  (T$_{\rm{spin}}$/c$_{\rm{f}}$) cm$^{-2}$ and N(\HI) =  3 $\times 10^{19}$ (T$_{\rm{spin}}$/c$_{\rm{f}}$) cm$^{-2}$ respectively. 
In Sec \ref{dichotomy} we expand on these results using stacking techniques. 

In Fig. \ref{fig:FluxDistr} we compare the flux density distribution of detections and non-detections using the Kolmogorov-Smirnov (K-S) test. The probability that the two distributions are different is only 10$\%$, implying that, statistically, detections and non-detections have a similar flux distribution. \HI\ detections are found down to the lowest fluxes in our sample. This is an important result, because it shows that our detections are not biased toward brighter sources and suggests that systematic \HI\ absorption studies can be carried out at even lower radio fluxes.

The detections in the $g - r > 0.7$ AGN sample reveal a wide variety of \HI\ profiles. In Fig. \ref{Dvmain} we show that we detect profiles where the peak of the absorption is redshifted up to +300 km s$^{-1}$, and blueshift up to $- 200$ km s$^{-1}$ compared to the systemic velocity. 
The analysis of the individually detected profiles shows that the median FWHM of the \HI\ main absorption components is $\sim$ 62 \kms. A detailed discussion of analysis of the complex profiles will be given in an upcoming paper presenting the full dataset. As a general conclusion, the diversity of the detected profiles suggests that the kinematics of \HI\ is quite complex in radio AGN.

\begin{figure}
\begin{center}
\includegraphics[width=.45\textwidth]{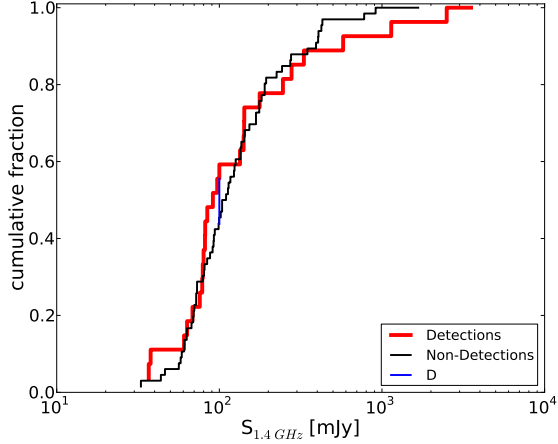}
\caption{Kolmogorov-Smirnov test for the flux distribution of detections (red line) and non-detections (black line). D is the maximum distance of the set of distances between the cumulative fraction of detections and non-detections. We measure $D = 0.117$ for the sample of 27 detections and 66 non-detections. }\label{fig:FluxDistr}
\end{center} 
\end{figure}

\begin{figure}[t!]
\begin{center}
\includegraphics[width=.45\textwidth]{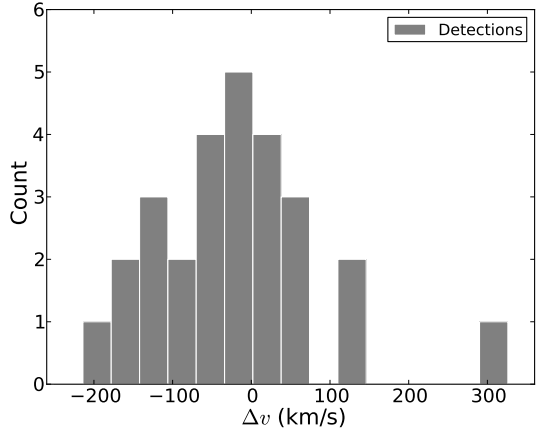}
\caption{Velocity offset of the \HI\ peak from the systemic velocity in the 27 detections.}\label{Width}\label{fig:WidthTest}\label{Dvmain}
\end{center}
\end{figure}

\subsection{\HI\ stacking: testing the \HI\ distribution of AGN down to low optical depth}\label{dichotomy}

One of the main goals of the study is to explore the presence of \HI\ at lower optical depth by using stacking techniques. We adapted the script used for \HI\ emission studies in \cite{Gereb} to perform stacking of \HI\ absorption. The spectra are extracted at the peak of the continuum source, assuming that the sources are unresolved. The stacking is done in optical depth, by aligning  the spectra based on their SDSS redshift. 

As shown in Fig. \ref{stack:DetectionNonDetection}, stacking a large number (66) of undetected sources results in an average non-detection with a relatively sensitive 3-$\sigma$ upper limit of $\tau$ $<$ 0.002 (solid line), while the sample of the detected sources gives an average $\tau_{\rm peak}$ = 0.02 (dashed line). From Gaussian fitting we measure FWHM = 203 $\pm$ 7 \kms\ in the profile of stacked detections. The integrated optical depth of this profile is relatively high, N(\HI) =  $(7.4 \pm 0.2)$ $\times$ 10$^{18}$ (T$_{\rm{spin}}$/c$_{\rm{f}}$) cm$^{-2}$. Using the median FWHM of $\sim$ 62 km s$^{-1}$ and the 3$-\sigma$ upper limit $\tau$  = 0.002, we derive N(\HI) $<$ $(2.26 \pm 0.06)$ $\times$ 10$^{17}$ (T$_{\rm{spin}}$/c$_{\rm{f}}$) cm$^{-2}$ for the upper limit of the stacked non-detections. In Fig. \ref{stack:S_NHI}, the red line marks the upper limit reached after stacking, showing the sensitivity improvement compared to single detections and upper limits. 

We have checked that the width of the co-added profiles is not affected by redshift inaccuracy during stacking in the following way. \cite{Maddox} show that for \HI\ emission, if the redshift errors are smaller than the median width of the co-added \HI\ profiles, the resulting width is not much affected by redshift inaccuracy. In our sample the 3-$\sigma$ SDSS redshift errors are lower than the median FWHM = 62 \kms\ (see Sec \ref{Sec:SampleSelection}), suggesting that the width of the stacked profile is not affected by redshift inaccuracy. However, \cite{Maddox} also show that the error distribution of the SDSS redshifts has broad, non-Gaussian tails to large values, which may have an effect on the stacking.

In Fig.\ \ref{Dvmain} we show that a significant fraction of our sources are offset from the systemic velocity. This suggests that the width of the stacked profile can be affected by the velocity offset distribution of the detections. In order to test the possible effect of this, instead of using the SDSS redshift, we stack the \HI\ detections by shifting the spectra with the frequency of the \HI\ peak from Fig.\ \ref{Dvmain}. The resulting profile is plotted in Fig. \ref{stack:DetectionNonDetection} with a dotted line, showing that both the width and the peak $\tau$ of the stacked profile change, however, as expected, the integrated optical depth remains the same. From Gaussian fitting we measure FWHM = 107 $\pm$ 4 \kms. The stacked FWHM is higher than the median {FWHM = 62 \kms} of the individual \HI\ detections (see Sec.\ \ref{Detection_HI}). However, as we mentioned earlier in Sec.\ \ref{Detection_HI}, only the FWHM of the main \HI\ component is estimated in the individual detections because of the complexity of the \HI\ profiles. Therefore, the larger stacked FWHM must be the result of additional \HI\ components. At the 3-$\sigma$ significance level, a redshifted feature is present in the stacked spectra at $\sim$ +250 \kms, however after inspecting the profiles we find that this feature is dominated by only a few (about 2) sources with complex \HI\ morphology.

\begin{figure}
\begin{center}
\includegraphics[width=.45\textwidth]{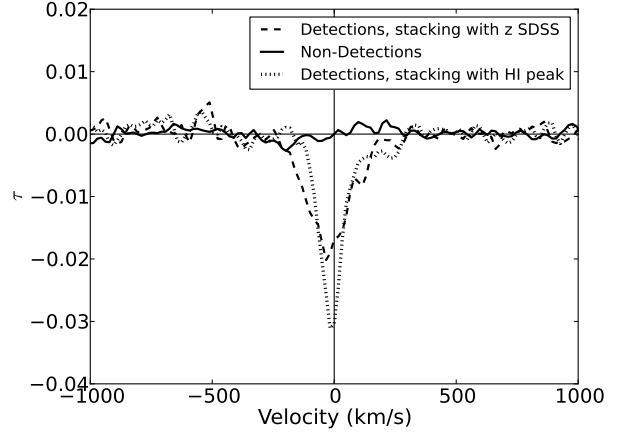}
\caption{The sample of 93 red AGN, the stacked profile of 27 detections (dashed and dotted lines, see explanation in the legend) and 66 non-detections (solid line) }\label{stack:DetectionNonDetection}
\end{center} 
\end{figure}

\begin{figure}
\begin{center}
\includegraphics[width=.45\textwidth]{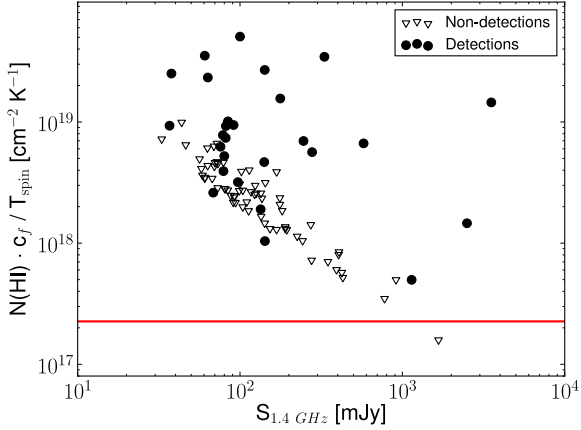}
\caption{The \HI\ column density versus the 1.4 GHz flux density for detections (filled circles)  and non-detections (empty triangles). The red line marks the N(\HI) upper limit of non-detections after stacking. }\label{stack:S_NHI} 
\end{center} 
\end{figure}

\begin{figure}
\begin{center}
\includegraphics[width=.45\textwidth]{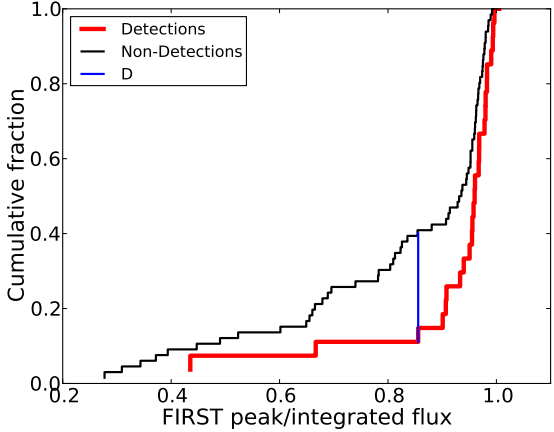}
\caption{Compactness distribution of detections and non-detections. We measure D = 0.319 for 27 detections and 66 non-detections. } \label{KS:compactness}
\end{center} 
\end{figure}

The interesting result from our stacking experiment is that the average \HI\ content of the non-detected sources is well below that of the detected sources, thus a strong dichotomy is found in the presence of \HI\ between detected sources and non-detections. The dichotomy is also present in Fig. \ref{stack:S_NHI}, where some sources have fairly high \HI\ integrated optical depths, however in many objects the N(\HI) is much lower. As already mentioned in Sec. \ref{Detection_HI} and shown in Fig \ref{fig:TauNHI}, the individually detected $\tau$ distribution in our sample is very broad, which is consistent with the broad range of \HI\ masses and column densities that is seen in \atlas. Hence, we cannot disentangle whether the dichotomy in the stacked profiles is due to a bimodal or a power-law distribution of the \HI\ column densities. This is, in fact, a long-standing problem that has been investigated for early-type galaxies for the past 30 years \citep{Knapp1985}.

Van Gorkom et al.\ (1989) - albeit using fewer objects and not using stacking techniques - already estimated the expected number distribution of $\tau$ in a sample of radio galaxies and concluded that, in general, only about 30 percent of the radio galaxies (with compact cores) have \HI\ with $\tau$ $>$ 0.01. Our result expands on this as we reach much lower values of $\tau$ after stacking. The fact that the stacked spectrum of those sources that are undetected individually does not show a detection implies that more sensitive observations would not yield many more new individual detections. Therefore we expect $\sim 30\%$ to be a representative detection rate for \HI\ in radio AGN.

One of the goals of our absorption stacking is to investigate the presence and signatures of blueshifted wings, indicative of outflows driven by jet-cloud  interactions. The peak optical depth of such absorption features is very low, below 1\% in most of the cases (see references from the Introduction). Therefore only in the brightest objects it is possible to reach such detection limits with current observations. Stacking makes it possible to decrease the optical depth limit and to test {whether the detection of outflows is a matter of sensitivity.}

\subsection{\HI\ and radio morphology}\label{Sec:HIRadioMorphology}

It has been suggested that different types of AGN, and in particular young radio sources, show different detection rates of \HI\ and to be surrounded by a different gaseous medium. Thus, we have separated our sample into compact and extended radio sources as described in Sec.\ 2. In Fig. \ref{KS:compactness}, the probability that the two data samples come from the same distribution is 2.5$\%$. This result implies that, statistically, detections and non-detections have a different compactness distribution. We find that compact sources (with $g - r > 0.7$) show a $\sim$ 42\% detection rate (20 detections, 28 non-detections), whereas only $\sim$16\% of extended sources are detected in \HI\ (7 detections, 38 non-detections).

From our compact sources 11 
are identified with a match in the COmpact RAdio sources at Low redshift (CORALZ) sample \citep{Snellen2004, deVries}, and VLBI observations assure us that CORALZ sources are intrinsically small (CSS and GPS), likely young AGN. For CORALZ sources we find an even higher detection rate of 55\%. 

The result of stacking in Fig. \ref{stack:CompactDetected} shows that the optical depth dichotomy of detections and upper limits also holds for compact and extended sources, in agreement with the dichotomy between detections and non-detections reported in Sec. \ref{dichotomy}. Furthermore, stacking reveals dissimilar profiles both regarding the peak optical depth and the width of the profiles. Gaussian fitting yields a FWHM of 203 $\pm$ 7 \kms\ for the stack of compact sources (similar to the total sample of detections), and 120 $\pm$ 13 \kms\ for the extended sources. The corresponding column density in compact sources is N(\HI) = $(8.5 \pm 0.3)$ $\times$ 10$^{18}$ (T$_{\rm{spin}}$/c$_{\rm{f}}$) cm$^{-2}$, whereas we detect lower columns of N(\HI) = $(2.9 \pm 0.2)$ $\times$ 10$^{18}$ (T$_{\rm{spin}}$/c$_{\rm{f}}$) cm$^{-2}$ for extended sources. 

\begin{figure*}[!t]
\begin{center}
\includegraphics[width=.45\textwidth]{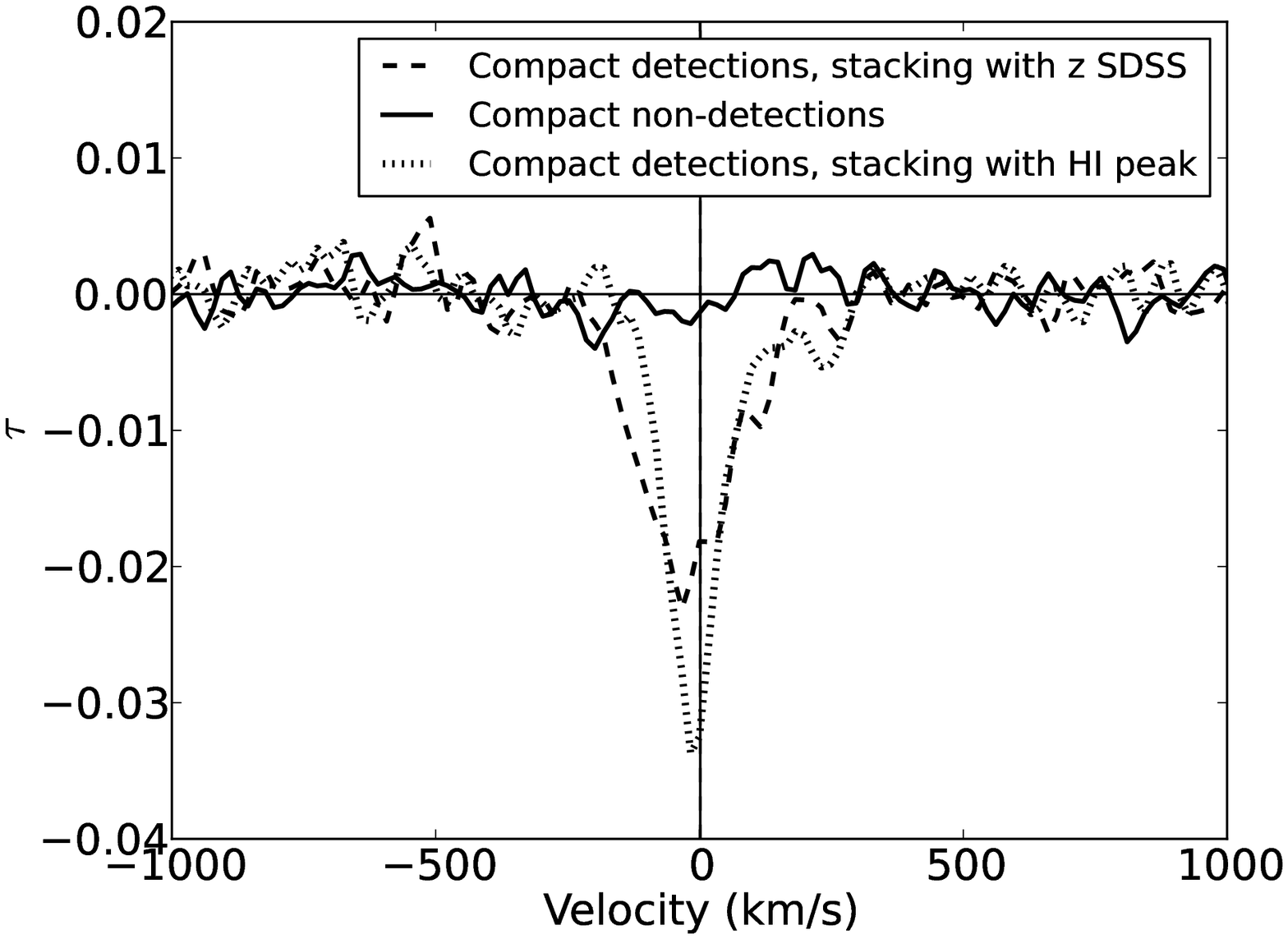}
\includegraphics[width=.45\textwidth]{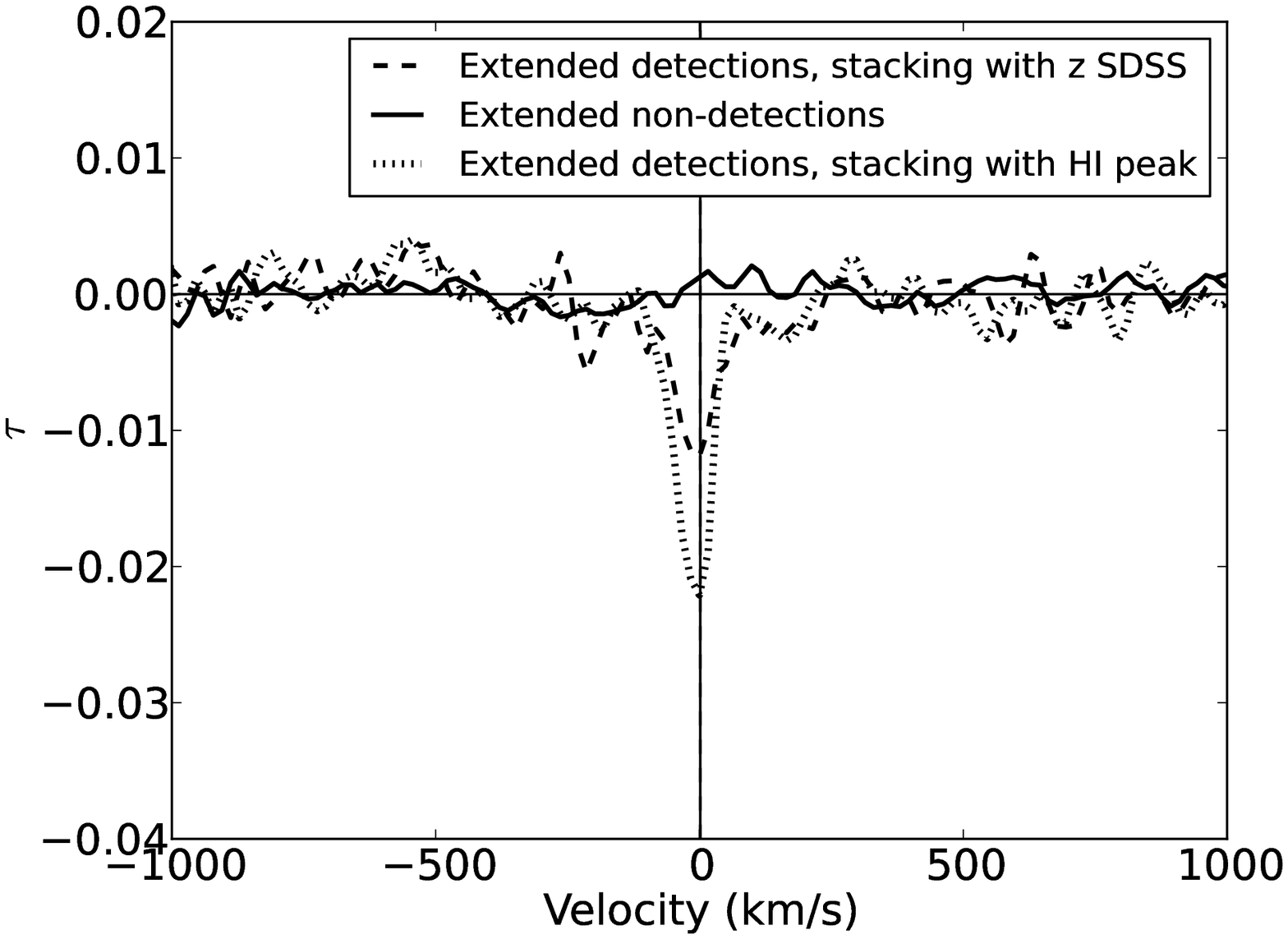}
\caption{The stacked profiles of compact (20 detections, 28 non-detections) and extended (7 detections, 38 non-detections) sources on the left and right side respectively. Detections are indicated by dashed and dotted lines like in Fig. \ref{stack:DetectionNonDetection}, and non-detections by solid lines.}\label{stack:CompactDetected}
\end{center} 
\end{figure*}

In order to test the effect of redshift differences on the measured profile widths, we repeat the stacking exercise from Sec.\ \ref{dichotomy} by shifting the spectra with the \HI\ peak. The stacked profiles become narrower, however the FWHM of the profiles of the compact and extended sources remains different. From Gaussian fitting we find FWHM = 115 $\pm$ 4 \kms\ for compact sources and FWHM = 75 $\pm$ 5 \kms\ for the extended sample. However, the integrated optical depth of compact sources is hardly affected, we measure N(\HI) = $(8.4 \pm 0.2)$ $\times$ 10$^{18}$ (T$_{\rm{spin}}$/c$_{\rm{f}}$) cm$^{-2}$. For extended sources the N(\HI) increases by 34$\%$ to N(\HI) = $(3.9 \pm 0.2)$ $\times$ 10$^{18}$ (T$_{\rm{spin}}$/c$_{\rm{f}}$) cm$^{-2}$. The difference in N(\HI) between compact and extended is still a factor of 2, meaning that our conclusions do not change.

From our WSRT observations we do not know the location of the absorption or the structure of the radio continuum. Higher resolution FIRST observations (5 arcsec angular resolution) are a better measure of the core brightness; therefore, we repeat the stacking exercise by deriving $\tau$ from the FIRST peak flux density (and by shifting the spectra with the optical redshift). For compact sources we measure N(\HI) = $(8.1 \pm 0.2)$ $\times$ 10$^{18}$ (T$_{\rm{spin}}$/c$_{\rm{f}}$) cm$^{-2}$ (5$\%$ change) and for extended sources N(\HI) = $(4.8 \pm 0.3)$ $\times$ 10$^{18}$  (T$_{\rm{spin}}$/c$_{\rm{f}}$) cm$^{-2}$ (65$\%$ change) in the stacked profiles. These differences are a result of the different resolution of our WSRT observations and the FIRST survey, and they are more prominent for the extended sources, as expected. However, even if all of the \HI\ absorption is against the core, the difference between the integrated optical depth of compact and extended sources is still a factor of $\sim$2.

The role of c$_{\rm{f}}$, i.e., \ the gas fraction covered by the radio source, can also be important for the interpretation of the observed optical depth. \cite{Curran2013} pointed out that the covering factor  (c$_{\rm{f}}$) is proportional to the $\tau_{\rm{obs}}$, and inversely proportional with the size of the radio source. Hence, a systematic difference in the covering factor of compact and extended sources could be influencing the measured optical depths.

In principle, the integrated optical depth difference between compact and extended sources in Fig. \ref{stack:CompactDetected} could be explained by a systematic difference in the covering factor. For example, only 30 percent of the continuum may be covered by the \HI\ screen (c$_{\rm{f}}$ $= 0.3$) in extended sources, while 100\% (c$_{\rm{f}}$ $= 1$) may be covered in compact sources.
VLBI studies available in the literature can be inspected to verify if such a trend exists. The covering factor is close to c$_{\rm{f}}$ = 1 in the compact source 1946+708 \citep{Peck}, however one can also find examples for compact sources with lower covering factor of c$_{\rm{f}}$ $\sim$ 0.2 in B2352+495 \citep{Araya} and in 4C 12.5 \citep{Morganti2013}. In extended sources c$_{\rm{f}}$ = 1 is estimated in the re-started source 3C 293 \citep{Beswick}. However lower covering factor of a few percent is measured in NGC 315 \citep{Morganti2009}, and c$_{\rm{f}}$ = 0.5 is estimated in the extended source 3C 305 \citep{Morganti2005}.
Thus,  from VLBI observations it is not at all clear that such a {\sl systematic} difference could realistically be present, as both compact  and extended sources show a broad range of covering factors. Thus, we conclude that at least part of the difference in optical depth between compact and extended sources is a real effect. However, VLBI observations of homogeneous AGN samples are needed to measure the c$_f$ and verify the difference between compact and extended sources at high resolution.

\subsection{The nature of the \HI\ absorbing systems}\label{Origin}

We can use the detailed studies of the \HI\ content of local early-type galaxies as a reference sample to investigate the nature of the \HI\ gas in our sample. \cite{Serra} detected \HI\ emission in about 40\% of \atlas\ early-type galaxies in the field. In 25\% of the total sample, i.e., in roughly half of the detected cases, the gas is distributed in a flattened disk/ring morphology. In the other half of the detected sample, \HI\ has an unsettled morphology.

In radio AGN the detection of \HI\ may strongly depend on the relative orientation of the \HI\ gas and the background continuum source. If some of the \HI\ structures have flattened morphology, the detection of \HI\ in these sources would be limited by orientation effects. To test this effect, we have inspected the inclination of our galaxies using the minor-to-major axis ratio in the $r$ band (expAB$\_r$), and using the optical images of the galaxies from the SDSS. The fact that the most highly inclined objects at $b/a < 0.6$ in Fig. \ref{Incl} all show \HI\ absorption suggests that orientation plays a role and that the \HI\ is likely in a flattened structure, such as a disk, in many sources. In the optical images we indeed see the presence of (nearly) edge-on disk morphologies in these objects. 

However, we do detect some galaxies at $b/a > 0.6$. \HI\ disks with low inclination are less likely to be detected, therefore the detection of \HI\ in the high $b/a$ regime implies that at least in some galaxies the \HI\ is not in a flattened structure, but it possibly has a more unsettled distribution. This is also supported by the fact that flattened structures in the $b/a < 0.6$ region all have relatively high integrated optical depth, while at higher $b/a$ detections span a wider N(\HI) range. 

Considering the 25\% detection rate of disks in \atlas\ galaxies, the 16\%  detection rate in the sample of extended sources is comparable to what one would expect to see if all detections of extended sources have a disk morphology. However, compact sources have a higher detection rate of 42\%, which is about the same as the total detection rate for \HI\ emission in the \atlas\ sample. Given that the detection of  \HI\ absorption requires a certain geometry (i.e.,\ the \HI\ has to be in front of the continuum source), this suggest that compact sources are at least as \HI-rich as the general population of early-type galaxies, but the total detection rate (\HI\ emission and absorption) could be even higher in compact sources.
The larger FWHM and higher \HI\ detection rate of compact sources suggests that their \HI\ gas often has more spherical morphology, less affected by orientation, e.g.,\ the \HI\ can be present in an unsettled distribution or thicker disks. Furthermore, the high integrated optical depths in compact sources, and in particular the young radio sources in our sample, indicates that young AGN are particularly rich in gas.
However, due to their small size, compact sources might have higher covering factors and therefore higher observed optical depths than extended sources \citep{Curran2013}. Hence, high-resolution observations are needed to estimate the effect of the covering factor on the measured optical depth. 

Our stacking exercise does not reveal a significant detection of outflowing gas at our current detection limit. As we will discuss in a forthcoming paper, about 10\% of the direct detections show a blueshifted wing, possibly due to an outflow with a typical peak optical depth of $\tau \sim  0.01$. The 10\% detection rate implies that we will need to reach at least a sensitivity of $\tau \sim  0.001$ in order to detect the contribution of such outflows in the stacked spectrum. This is below our detection limit and, therefore, it may not be too surprising that we do not see a blue wing in the stacked profile. Upcoming surveys with the SKA Pathfinders will result in a few hundred thousand new \HI\ detections, most of which will be above z = 0.1. Radio AGN are more numerous in the distant Universe, therefore we can expect to have new detections of \HI\ absorbers associated with radio continuum sources. Furthermore, surveys specifically designed for \HI\ absorption search, e.g., the ASKAP FLASH (The First Large Absorption Survey in \HI, \citealt{Johnston2008}), will deliver several hundred new detections, allowing one to reach the necessary sensitivity to detect, statistically, the jet-driven \HI\ outflows.

\begin{figure}[t!]
\begin{center}
\includegraphics[width=.45\textwidth]{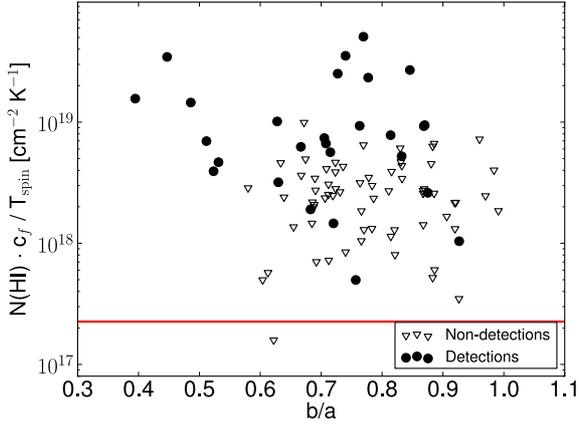}
\caption{Column density vs. the minor-to-major axis ratio from the exponential fit in $r$ band for detections and non-detections. The red line marks the N(\HI) upper limit of non-detections after stacking.}\label{Incl}
\end{center}
\end{figure}

\section{Conclusions and implications for the future}\label{Conclusions}

In this paper we carried out a systematic study of the \HI\ properties of radio AGN. Despite our shallow observations, we obtain a detection rate of 30 percent, similar to deeper studies. Our sample is larger than previous studies, allowing us to carry out \HI\ absorption stacking and to confirm that in the low-$\tau$ limit 30 percent is a representative \HI\ detection rate for the general population of radio AGN. We find a dichotomy in the presence of \HI\ in the sense that even when a large number of spectra are averaged, galaxies that do not show \HI\ absorption in their individual spectra remain undetected to an average column density of N(\HI) $<$ $(2.26 \pm 0.06)$ $\times$ 10$^{17}$ (T$_{\rm{spin}}$/c$_{\rm{f}}$) cm$^{-2}$. We argue that in many galaxies, the \HI\ must be in a flattened structure so that orientation effects play a role. 
This result is in good agreement with \cite{Curran2010}, who suggest that galactic disks are the bulk of \HI\ absorption in all types of AGN.
However, orientation effects alone cannot fully explain the dichotomy that we see in our sample, suggesting that some fraction of our galaxies must be depleted of cold gas.

Upcoming surveys will observe AGN over a large flux density range. Our results suggest that the detection rate does not depend on the apparent flux of a source and this has positive implications for future, deeper surveys. These large-area surveys will uncover a very large number of \HI\ absorptions systems. According to the 1.4 GHz radio luminosity function, star forming galaxies become dominant at radio power lower than $<$ 10$^{23}$ W/Hz \citep{Mauch2007}, therefore at lower fluxes we expect a mix of star-forming/AGN populations.

The detection of \HI\ absorption depends strongly on the strength of the underlying continuum; therefore, it is possible to detect absorption up to high redshift, if the continuum is strong enough. Unlike \HI\ absorption, \HI\ emission studies are limited by sensitivity in the higher redshift Universe. However, if emission and absorption are tracing similar morphological structures, \HI\ absorption studies can be used just as efficiently to find cold gas not just in AGN, but also in star-forming galaxies at higher redshift. The increased number of sources will provide enough data to perform \HI\ stacking experiments and, hence, to probe the highest redshift regime of the observed radio sky at low optical depth. Thus, even though \HI\ absorption only traces the cold gas component with T$_{\rm{spin}}$ up to a few $\times$ 1000 K, it can still provide important information (including redshift) for the detected objects. 

Compact sources show higher detection rates, optical depths and FWHM than extended sources, strongly suggesting that different gas conditions exist in these two types of radio sources; however, high resolution observations and a better measure of the covering factor are needed to confirm this result. It seems that a large fraction of compact sources reside in a gas rich environment, and the nuclear activity in most of the young AGN is connected with the presence of unsettled gas.

In a forthcoming paper we will publish the details of our \HI\ detections and the AGN sample. We will explore in more detail the parameter space of our \HI\ profiles  (e.g., blueshift/redshift, width, asymmetries) in relation with radio source and host galaxy properties. We aim to understand if (and to what extent) the properties of the detections are related to the strength and morphology of the radio AGN.

\section{Acknowledgements}
The WSRT is operated by the ASTRON (Netherlands Foundation for Research in Astronomy) with support from the Netherlands Foundation
for Scientific Research (NWO).

RM gratefully acknowledge support from the European Research Council under the European Union's Seventh Framework Programme (FP/2007-2013) /ERC Advanced Grant RADIOLIFE-320745.

\end{document}